# Bianchi V Brane-world Cosmology with a Generalized Chaplygin Gas

Ö. Sevinç and G. Berköz

*Department of Physics, Science Faculty, Istanbul University*

**Abstract.** In the context of brane cosmology, we investigate exact solution of the gravitational field equations in the Randall-Sundrum model for an anisotropic brane with Bianchi V geometry, with a generalized Chaplygin gas, which plays the role of dark energy and interacts with dark matter particles. Also, the behavior of the observationally important parameters like expansion and anisotropy is examined.

**Keywords:** Brane-world, Dark Matter-Dark-Energy, Chaplygin Gas.

**PACS:** 04.50.+h, 98.80.Jk, 95.30.Cq and 95.35.+d.

## INTRODUCTION

In order to fully account for the existing observations [1], one must bring in at least two additional new mysteries: the concept of Dark Matter [2], and the idea of a smoothly distributed energy that cannot be identified with any form of matter, the so-called Dark Energy [3]. Though there are no direct laboratory observational or experimental evidence for both of them yet, a unified dark matter-dark energy scenario i.e., they are two different manifestations of a single fluid would be interesting. Recently, this unified model has been proposed and known as generalized Chaplygin gas having exotic equation of state as follows

$$p = -B/\rho^\alpha \qquad (1)$$

where B is a positive constant and α is a constant in the range $0<\alpha\leq 1$ [4]. In our present work, we consider a more general Modified Chaplygin Gas (MCG) obeying an equation of state

$$p = \gamma\rho - B/\rho^\alpha \qquad (2)$$

where $0\leq \gamma \leq 1$, $0\leq \alpha \leq 1$ and $B>0$ [5].

Localizing matter field and gravity in a submanifold has been utilized recently by Randall and Sundrum for non-factorizable geometries in five dimensions [6]. They have shown for this five dimensional geometry that it is possible to confine a single massless bound state in a domain wall or three-brane. Hence all matter and gauge fields (except gravity) are confined in a three-brane embedded in a five dimensional space-time [7]. Gravity on the brane can be described as follows

$$G_{\mu\nu} = -\Lambda g_{\mu\nu} + k_4^2 T_{\mu\nu} + k_5^4 S_{\mu\nu} - E_{\mu\nu} \qquad (3)$$

where $S_{\mu\nu}$ is the local quadratic energy-momentum correction and $E_{\mu\nu}$ is the nonlocal effect from the bulk free gravitational field:

$$E_{\mu\nu} = \left(\frac{k_5}{k_4}\right)^4 \left[U(u_\mu u_\nu + \frac{1}{3}h_{\mu\nu}) + P_{\mu\nu} + 2Q_{(\mu}u_{\nu)}\right] \qquad (4)$$

where $U$ is a scalar, $Q_\mu$ a spatial vector and $P_{\mu\nu}$ a spatial, symmetric and trace-free tensor [8]. We assume that cosmological fluid on the brane is perfect, without dissipative effects. Then the matter correction terms are given by

$$S_{\mu\nu} = \frac{1}{12}\rho^2 u_\mu u_\nu + \frac{1}{12}\rho(\rho+2p)h_{\mu\nu} \qquad (5)$$

It is the aim of this work to investigate some classes of exact solutions of the gravitational field equation in the brane world model for the anisotropic Bianchi type V model, by assuming that the matter source on the brane consists of the MCG.

## GENERAL SOLUTION

The Einstein gravitational field equation for the Bianchi type V brane filled with a cosmological fluid obeying the MCG take the form:

$$\dot{\rho} + 3H(p+\rho) = 0 \qquad (6)$$

$$\frac{1}{V}\frac{d}{dt}(VH_i) - \frac{2a_1^2}{R_1^2} = \Lambda - \frac{k_4^2}{2}\left\{(\gamma-1)\rho - \frac{B}{\rho^\alpha}\right\} - \frac{k_4^2}{2\lambda}\left(\gamma\rho^2 - \frac{B}{\rho^{\alpha-1}}\right) + \frac{2U}{k_4^2\lambda} \qquad (7)$$

$$3\dot{H} + \sum_{i=1}^{3} H_i^2 = \Lambda - \frac{k_4^2}{2}\left\{\rho(1+3\gamma) - \frac{3B}{\rho^\alpha}\right\} + \frac{6U}{\lambda k_4^4} + \frac{k_4^2}{2\lambda}\left\{\rho^2(3\gamma+2) - \frac{3B}{\rho^{\alpha-1}}\right\} \qquad (8)$$

Using equation (2) and (6), we find energy density of the MCG as follows:

$$\rho = \left[\frac{B}{1+\gamma} + \frac{C}{V^{(1+\gamma)(1+\alpha)}}\right]^{\frac{1}{1+\alpha}} \qquad (9)$$

where $V$ is the volume scale factor and C is an arbitrary constant of integration. If we define a new function $\ddot{V} = F(V)$, and use (7), we can easily obtain general solution in the form:

$$t - t_0 = \int \frac{dV}{\sqrt{2\int F(V)dV + C_1}} \qquad (10)$$

where $C_1$ and $t_0$ are constants of integration. Thus, the general solution of the gravitational field equation can be expressed in the following exact parametric forms

$$\theta = \frac{\sqrt{2\int F(V)dV + C_1}}{V} \qquad (11)$$

$$A = \frac{3K^2}{2\int F(V)dV + C_1} \qquad (12)$$

By introducing two new variable $\tau$ and $v$ by means of the transformation $V = V_0 v$ and $\tau = \sqrt{3}k_4 B^{1/2(1+\alpha)}t$ with $V_0$ = constant, by normalizing the value of the arbitrary integration constants so that $4a_1^2/k_4^2 B^{1/1+\alpha} = 1$, $C/BV_0^{(1+\gamma)(1+\alpha)} = 1$, $4U_0 V_0^{-4/3}/k_4^4 B^{1/(1+\alpha)}\lambda = 1$, $C_1/3V_0^2 k_4^2 B^{1/(1+\alpha)} = 1$, $B^{1/(1+\alpha)}/\lambda = b$, $2\Lambda/3k_4^2 B^{1/(1+\alpha)} = \lambda_0$ and $f(v) = 1/(1+\gamma) + 1/v^{(1+\gamma)(1+\alpha)}$, it follows that Eq.(10) can be written in the form:

$$\Delta\tau = \int \frac{dv}{\sqrt{\int\{v^{1/3} + \lambda_0 v - v[(\gamma-1)f^{1/(1+\alpha)} - f^{-\alpha/(1+\alpha)}] - bv[\gamma f^{2/(1+\alpha)} - f^{(1-\alpha)/(1+\alpha)}] + v^{-1/3}\}dv + 1}} \tag{13}$$

where $\Delta\tau = \tau - \tau_0$.

## CONCLUSIONS

Using eq. (10), (11) and (13), the variation of the expansion scalar θ as a function of the dimensionless time parameter $\tau$ is represented in Fig. 1. Evolution of another observationally important parameter which is called anisotropy is drawn in Fig. 2. The behavior of this parameter in the brane world model for the anisotropic Bianchi type V geometry is totally different from the standard cosmology [9]. Also, the behaviors of anisotropy and expansion parameters in Bianchi type V and I have approximately same characters [10].

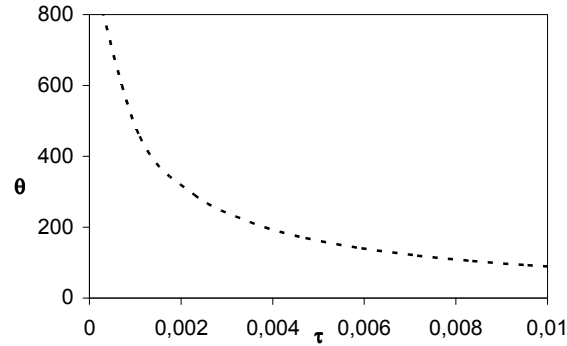

**FIGURE. 1**: The behavior of expansion scalar θ for $\gamma$ =1/3, α=1/2 and b=0.01.

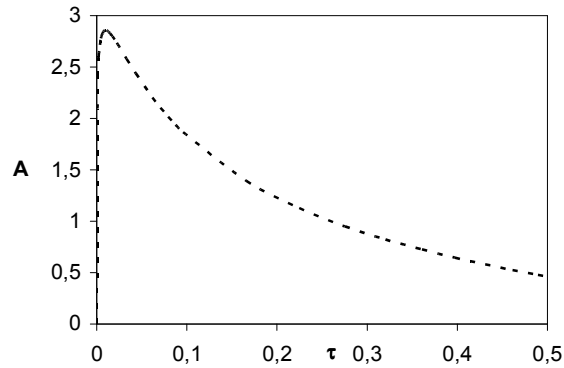

**FIGURE 2**: Anisotropy parameter A as a function of the dimensionless time parameter $\tau$ for $\gamma$ =1/3, α=1/2 and b=0.01.

## REFERENCES
1. Riess, A. G. et. al. Astron. J. **116**, 1009(1998).
    Bennett, C. L. et. al. Astrophys. J. Suppl. Ser. **148**, 1 (2003).
2. Navarro, J.F. et. al. The Astrophysical J. 462-563(1996).
3. S. M.Carroll, Living Rev. Rel. **4**, 1 (2001).
4. M. C. Bento, Physical Rev. D, **67**, 063003 (2003).
5. U. Debnath, A. Banerjee and S. Chakraborty, Class. Quantum Grav. **21**, 5609-5617 (2004)
6. L. Randall, and Sundrum, R, Physical Review Letters, **83**, 4690 (1999).
7. P. Brax, and Van de Bruck, C. Classical and Quantum Gravity, **20**, R 201 (2003).
8. M. Sasaki, et. al. Phys. Rev. D, **62**, 024008 (2000).

9. E. Güdekli, "Horizon-Isotropization Problems in Bianchi Type Solution", Ph.D. Thesis, Istanbul University, 2004.
10. Mak, M. K. and T. Harko, Physical Rev. D, **71**, 104022 (2005).